\pdfoutput=1

\documentclass[prl,twocolumn,superscriptaddress,letterpaper]{revtex4}

\usepackage{graphicx,float}
\usepackage{epsfig}
\usepackage{dcolumn}
\usepackage{bm,mathptmx}
\usepackage{amsmath}
\usepackage{amssymb}
\usepackage[colorlinks=true,citecolor=blue,urlcolor=blue]{hyperref}

\newcommand{\ket}[1]{\ensuremath{|\,{#1}\,\rangle}}
\newcommand{\bra}[1]{\ensuremath{\langle\,{#1}\,|}}

\begin{document}


\title{Applying the Simplest Kochen-Specker Set for Quantum Information Processing}


\author{Gustavo~Ca\~nas}
\author{Mauricio~Arias}
\author{Sebasti\'an~Etcheverry}
\author{Esteban~S.~G\'omez}
\affiliation{Departamento de F\'{\i}sica, Universidad de Concepci\'on, 160-C Concepci\'on, Chile}
\affiliation{Center for Optics and Photonics, Universidad de Concepci\'on, Concepci\'on, Chile}
\affiliation{MSI-Nucleus for Advanced Optics, Universidad de Concepci\'on, Concepci\'on, Chile}

\author{Ad\'an~Cabello}
\affiliation{Departamento de F\'{\i}sica Aplicada II, Universidad de Sevilla, E-41012, Sevilla, Spain}

\author{Guilherme~B.~Xavier}
\affiliation{Center for Optics and Photonics, Universidad de Concepci\'on, Concepci\'on, Chile}
\affiliation{MSI-Nucleus for Advanced Optics, Universidad de Concepci\'on, Concepci\'on, Chile}
\affiliation{Departamento de Ingenier\'{\i}a El\'ectrica, Universidad de Concepci\'on, 160-C Concepci\'on, Chile}

\author{Gustavo~Lima}
\affiliation{Departamento de F\'{\i}sica, Universidad de Concepci\'on, 160-C Concepci\'on, Chile}
\affiliation{Center for Optics and Photonics, Universidad de Concepci\'on, Concepci\'on, Chile}
\affiliation{MSI-Nucleus for Advanced Optics, Universidad de Concepci\'on, Concepci\'on, Chile}


\begin{abstract}
Kochen-Specker (KS) sets are key tools for proving some fundamental results in quantum theory and also have potential applications in quantum information processing. However, so far, their intrinsic complexity has prevented experimentalists from using them for any application. The KS set requiring the smallest number of contexts has been recently found. Relying on this simple KS set, here we report an input state-independent experimental technique to certify whether a set of measurements is actually accessing a preestablished quantum six-dimensional space encoded in the transverse momentum of single photons.
\end{abstract}


\date{\today}

\maketitle


{\em Introduction.---}For any quantum system of a given dimension $d \ge 3$, there is always a set of yes-no tests for which, no matter how the system is prepared, the predictions of quantum theory cannot be reproduced with any theory that assumes that the measurement results are predefined and independent of other compatible measurements \cite{Specker60,KS67}. These sets, called Kochen-Specker (KS) sets \cite{PMMM05}, provide a proof of the impossibility of explaining quantum theory with noncontextual hidden variables \cite{Specker60,KS67} and are also important for other fundamental results in quantum theory \cite{HR83,CK06,CK09,Cabello08,BBCP09,KZGKGCBR09,ARBC09,DHANBSC13,AGACVMC12}. In addition, KS sets have potential applications in quantum information processing, since they are essential for nonlocal games \cite{RW04}, games with quantum state-independent advantage \cite{RW04,SS12,DHANBSC13}, quantum key distribution secure against attacks based on classical simulations of complementarity \cite{Svozil10,CDNS11}, single-shot entanglement-assisted zero-error communication \cite{CLMW10,MSS13}, and state-independent dimension witnessing with sequential measurements \cite{GBCKL13}. However, so far, the intrinsic complexity of KS sets has prevented the experimental development in this direction.

The original set found by Kochen and Specker consists of 117 tests on a quantum system of dimension $d=3$ that can be grouped in 132 contexts (i.e., sets of mutually jointly measurable tests) \cite{KS67}. The smaller the number of contexts a KS set has, the easier it is to observe the contrast between the predictions of quantum theory and those of noncontextual hidden variables. In this sense, the discovery of simpler KS sets \cite{Kernaghan94,KP95,CEG96} has allowed the first experimental investigations \cite{DHANBSC13,Canas13}.

Remarkably, the KS set with the smallest number of contexts has been found only very recently. It consists of a set of 21 tests on a quantum system of dimension $d=6$ that can be grouped in seven contexts \cite{LBPC13}. In practical terms, this set (hereafter called KS21) provides a shortcut for applying KS sets for quantum information processing.

Here we show how KS sets, and specifically KS21, can be used to test whether a set of measurements is actually accessing a preestablished $d$-dimensional quantum system (e.g., the six-dimensional quantum system defined by the six lowest energy levels of the Er$^{3+}$ ion). In particular, the problem we address is the following: Bob receives from Alice preestablished $d$-dimensional quantum systems prepared in an unknown state, possibly noisy. Bob has to check whether his measurements actually access Alice's $d$-dimensional quantum systems. In particular, Bob wants to be sure that his results cannot be produced by measurements on classical systems or different quantum systems (e.g., six energy levels of a different ion). We therefore show a direct application of KS sets: the certification of measurement hardware for high-dimensional quantum information processing.

These KS-based measurement dimension witnesses complement the device-independent dimension witnesses (DI-DWs) introduced in Ref.~\cite{GBHA10} and experimentally implemented recently in Refs.~\cite{Nat12a,Nat12b,DBSBLC14}. KS-based measurement dimension witnesses differ from DI-DWs in many senses: DI-DWs' purpose is to assess the minimum, classical or quantum, dimension that a set of preparations actually produce. DI-DWs do not make any assumption on the preparation and measuring devices' inner workings, but cannot distinguish preparations of quantum systems of dimension $d$ from preparations of classical systems of dimension larger than $d$, and require Alice to send, at least, $d+1$ different preparations.

The protocol for assessing whether Bob's set of measurements is accessing Alice's $d$-dimensional quantum system is the following. Bob starts by building measurement devices to test, on the quantum systems that he thinks Alice will send, each of the yes-no tests $\Pi_i$ of a $d$-dimensional KS set. Then, he checks that his devices produce results that satisfy the relations of pairwise mutual exclusivity of the KS set. That is, Bob checks that, for any pair of mutually exclusive tests $(\Pi_i,\Pi_j)$, his measurements satisfy that, if the system is in the state corresponding to the yes result for $\Pi_i$, then the result of $\Pi_j$ is always no, and vice versa. Notice that these relations of exclusivity may also be produced with measurements on classical or different quantum systems of dimension $d$ or higher.

To certify that his measurements access Alice's $d$-dimensional quantum systems, when he receives Alice's systems, Bob measures on them the frequency with which each $\Pi_i$ gives result yes. This allows him to reach a conclusion based on the following results: (i) The relations of exclusivity of a $d$-dimensional KS set can only occur with tests on classical or quantum systems of dimension $d$ or higher. (ii) For any $d$-dimensional KS set, there is a noncontextuality (NC) inequality violated by any $d$-dimensional quantum state by the same amount \cite{BBCP09}. (iii) The noncontextual bound and the maximum quantum value of any NC inequality are, respectively, given by the independence number $\alpha(G)$ and the Lov\'asz number $\vartheta(G)$ of the NC inequality's exclusivity graph $G$. This graph is defined as the one in which, when the correlations are expressed as a positive combination $\Sigma$ of probabilities, tests are represented by vertices and mutually exclusive tests by adjacent vertices \cite{CSW10,CSW14}. Therefore, if Bob has confirmed the relations of the exclusivity, the experimental value of $\Sigma$ should be $\vartheta(G)$ for any quantum state Alice may have prepared, but only if Bob's measurements are accessing Alice's $d$-dimensional quantum system. Otherwise, the experimental value for $\Sigma$ would be smaller than or equal to $\alpha(G)$, even in the case Bob's measurements are accessing a quantum system, but not the right one. For example, if Alice is encoding $d$-dimensional quantum information in the transverse momentum state of photons of a certain wavelength, but Bob's measurements do not work properly for that wavelength, then Bob will observe a value smaller than or equal to $\alpha(G)$.

In this Letter we exploit the simplicity of KS21 to experimentally demonstrate the usefulness of KS sets to certify that our measurements are actually accessing the quantum six-dimensional space encoded in the transverse momentum state of single photons. For that, we will use the following NC inequality:
\begin{equation}
 \label{IneKS21}
 \Sigma= 2 \times \sum_{i=1}^{21} P(\Pi_i=1) \stackrel{\mbox{\tiny{ NCHV}}}{\leq} 6,
 \end{equation}
where $P(\Pi_i=1)$ is the probability of obtaining result~1 (yes) when performing the test $\Pi_i=|v_i\rangle \langle v_i|$, where $|v_i\rangle$ with $i=1,\ldots,21$ are the KS states introduced in Ref.~\cite{LBPC13}. $\stackrel{\mbox{\tiny{ NCHV}}}{\leq} 6$ indicates that the upper bound for $\Sigma$ is $6$ for any noncontextual hidden variable theory. In contrast, in quantum theory the value of $\Sigma$ is $7$, regardless of the quantum state of Alice's six-dimensional system.


{\em Description of the experimental setup.---}In order to encode six-dimensional quantum information we employ the linear transverse momentum of single photons. In this case, a six-dimensional quantum state is created by defining six path possibilities for the photon transmission through a diffractive aperture. To produce each of the 21 KS states of KS21, we use a set of six parallel slits dynamically generated using a sequence of two spatial light modulators (SLM). SLMs are optical elements usually composed of a liquid crystal display matrix and linear polarizers \cite{Twi1,Twi2}. For this experiment, we further optimized the configuration of the SLM by resorting to quarter wave plates (QWP) placed between the linear polarizers. This was done to cover the phase modulations required for the generation of the KS21 states which have complex components. In fact, KS21 is the only critical (i.e., that does not contain simpler KS sets) $d$-dimensional KS set known that cannot be implemented in a $d$-dimensional real Hilbert space.

If the transverse coherence length of the beam is larger than the distance separating the first and the sixth slit, the state of the transmitted photon is given by \cite{Neves05,Lima09}
\begin{equation}
\ket{\psi} = \frac{1}{\sqrt{C}}\sum_{l=-l_6}^{l_6} \sqrt{t_{l}}e^{i\phi_l}\ket{l},
\label{State}
\end{equation}
where $l_6=5/2$ and $\ket{l}$ represents the state of the photon transmitted by the \textit{l}th slit \cite{Neves05}. $t_l$ ($\phi_l$) is the transmissivity (phase) defined for each slit and $C$ the normalization constant. Our slits are $64$~$\mu$m wide and have a separation between them of $128$~$\mu$m. The advantage of using SLMs to define slits is that they allow us to control $t_l$ and $\phi_l$, independently for each slit modulated. The versatility of SLMs has been proven to be crucial for fundamental investigations of quantum information with high-dimensional systems \cite{Nat12a,Lima09,SPadua_opex,Padgett_nature_phys,Lima11,Neves13,Lima13,Sciarrino_srep_2013}.

\begin{figure}[t]
\centering
\includegraphics[width=0.45 \textwidth]{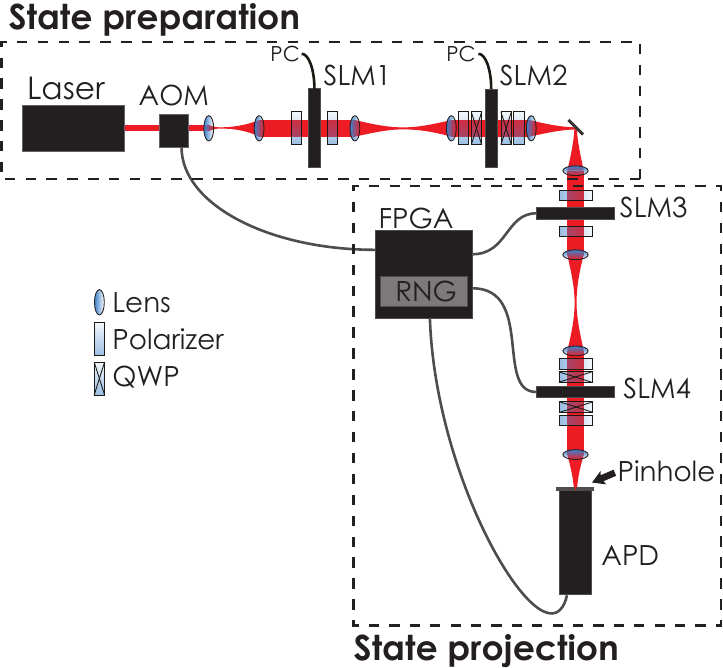}
\caption{(color online) Experimental setup. The state preparation stage consists of a attenuated single-photon source and of two SLMs used to encode one of the 21 KS states on the linear transverse momentum of a single photon. These SLMs are connected to a personal computer (PC) which, for clarity's sake, is not shown in the figure. The measurement projection stage employs two other SLMs to randomly project the incoming state onto one of the 21~KS states. An FPGA unit controls the entire experimental setup for testing the NC inequality. See main text for details. \label{fig1}}
\end{figure}

The experimental setup is depicted in Fig.~\ref{fig1}. It consists of two main blocks, the initial state preparation and the projection stages. The single photons used in the experiment are actually weak coherent states produced from heavily attenuated optical pulses. For the pulse generation, an acousto-optic modulator (AOM) is placed at the output of a continuous-wave diode laser at 690~nm. The weak coherent states are finally generated with an optical attenuator. As we mentioned, the amplitude and phase modulations required for the generation of the 21~KS vectors are obtained with a combination of two SLMs, SLM1 and SLM2. SLM1 controls the real part of the coefficients of the generated states, while SLM2 their phases \cite{Lima13}. The SLM2 is located at the image plane of the first one.


The generated state is then propagated through an imaging telescoping set of lens to the projection stage. The function of this stage is to project the transmitted state, onto any of the 21~KS set states, allowing us to implement the corresponding 21 yes-no questions. Two SLMs are again used in the same configuration used in the state preparation stage, SLM3 for amplitude and SLM4 for phase modulation. A quantum random number generator (RNG) IDQUANTIQUE QUANTIS is used to randomly choose one of 21 measurement projections to be applied. The projection is concluded after the output light from SLM4 is focused with a lens and detected by a pointlike avalanche single-photon detector (APD) placed at the center of the focal plane \cite{Lima11,Lima10,Leo10}. The pointlike detector is built with a 10~$\mu$m diameter pinhole placed in front of the APD. In this configuration, the probability of a single-photon detection is proportional to $|\bra{\psi}\text{KS}_i\rangle|^2$ \cite{Lima11}, with $i=1,\ldots,21$ and where $\ket{\psi}$ is the prepared state and $\ket{\text{KS}_i}$ is the KS state onto which the measurement projects.

The entire experimental setup is actively controlled by field-programmable gate array (FPGA) electronics in order to automate the measurements. A single FPGA unit is used to send trigger pulses to the AOM at a repetition rate of 30 Hz in order to create the faint optical pulses. The same FPGA is also used to synchronously change the masks in SLM3 and SLM4 for each pulse, and to record whether or not a detection occurred for this pulse. The unit then keeps track of all the detections for each combination of $\ket{\psi}$ and $\ket{\text{KS}_i}$. From the statistics of the recorded results one can obtain the value of $\Sigma$ appearing in the inequality (\ref{IneKS21}).


\begin{figure}[t]
\centering
\includegraphics[width=0.46 \textwidth]{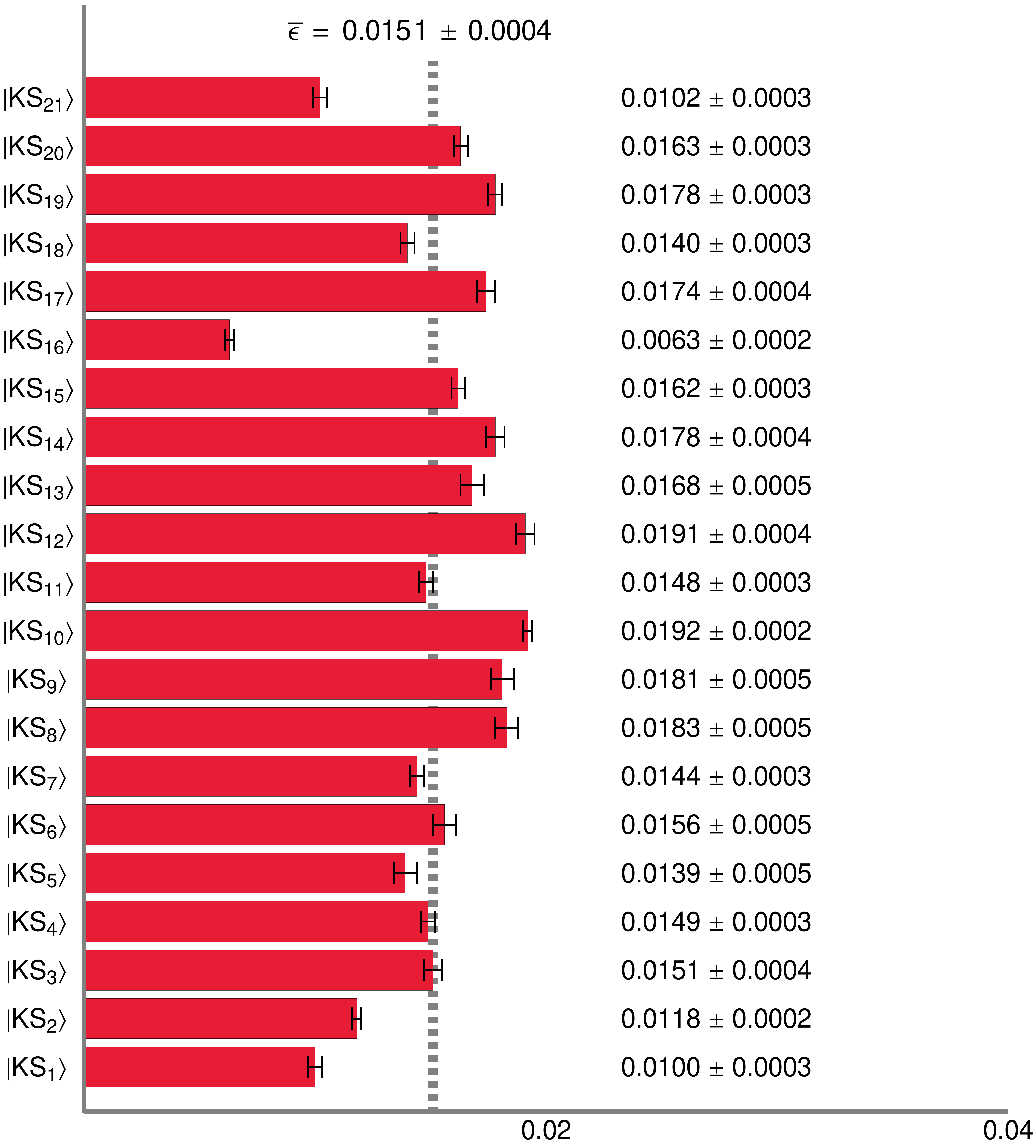}
\caption{(color online) Exclusivity tests performed for the 21 KS states. In these tests, the prepared state and the measured states are KS states that are supposed to be orthogonal. The KS states are numbered as they appear in Eq.~(1) in Ref.~\cite{LBPC13}. The dashed line shows the parameter $\overline{\epsilon}$ defined in the main text. \label{fig2}}
\end{figure}


{\em Methods and results.---}The noncontextual bound of the NC inequality (\ref{IneKS21}) is derived under the assumption that the states have certain orthogonality relations and, in particular, that the probability of obtaining a yes answer when the system is prepared in an orthogonal state is zero. However, under real experimental conditions, it is expected that the measured probabilities corresponding to these orthogonal projections will not be exactly zero. This originates from imperfections inherent to any experiment. These imperfections must be taken into account to properly correct the noncontextual limit of inequality (\ref{IneKS21}). As explained in Ref.~\cite{DHANBSC13}, this can be done by testing which is the fraction of experiments in which a ``wrong'' result is observed. In our case, we obtained that this fraction is, after averaging over all the orthogonalities in the KS set, $\overline{\epsilon} = 0.0151 \pm 0.0004$. Figure.~\ref{fig2} shows how this fraction differs for the orthogonalities of each state $\ket{\text{KS}_i}$ (in KS21 all states have the same number of orthogonalities). Since $\Sigma$ is the sum of 21 probabilities, one can easily calculate the corrected noncontextual bound assuming that the bound of the original inequality (\ref{IneKS21}) is only valid for a fraction $(1-\overline{\epsilon})$ of the experiments, and assuming the worst-case value for the other fraction. This gives an upper bound for $\Sigma$ equal to $6(1-\overline{\epsilon}) + 42\overline{\epsilon} = 6.55$ \cite{DHANBSC13}, which is still lower than the quantum value for the case of an ideal experiment, which is $7$. Notice that the same reasoning applies to the quantum value for an ideal case, yielding upper and lower bounds to it. This procedure is crucial to correctly estimate the expected result of the experimental certification of whether the hardware under test is accessing the correct $d$-dimensional quantum system.

\begin{figure}[t]
\centering
\includegraphics[width=0.46 \textwidth]{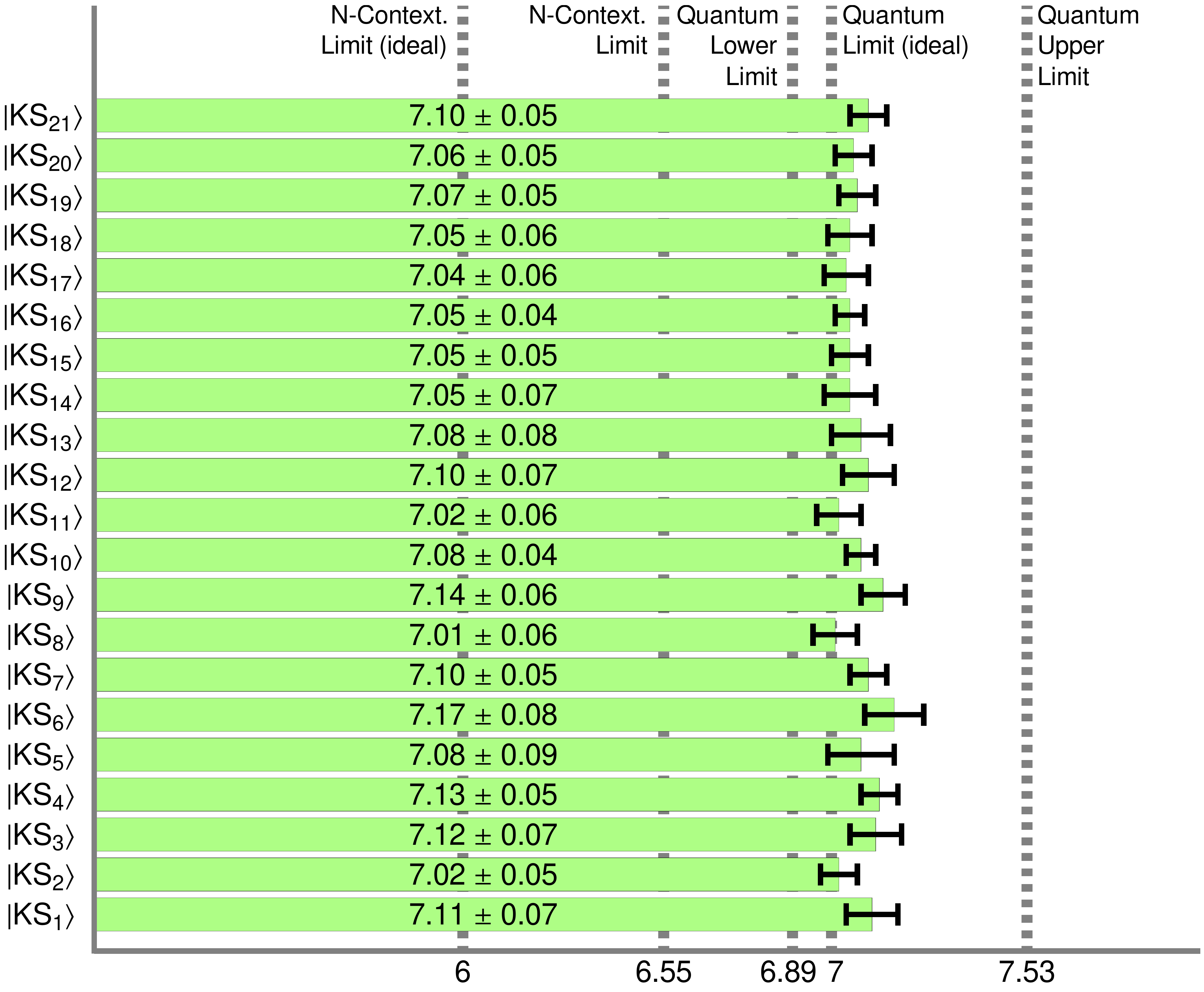}
\caption{(color online) Experimental results for $\Sigma$, defined in Eq.~(\ref{IneKS21}), using as initial states each of the KS states of KS21. The NC inequality (\ref{IneKS21}) is equally violated for all the states tested. $N$-Context Limit (Ideal) indicates the noncontextual bound in the ideal case in which $\overline{\epsilon}=0$. $N$-Context Limit indicates the bound when the actual value of $\overline{\epsilon}$ obtained in the exclusivity tests is taking into account. Quantum Limit (Ideal) indicates the expected quantum result when measurements access the correct $d$-dimensional quantum system in the ideal case in which $\overline{\epsilon}=0$. Quantum Lower (Upper) Limit indicates the limits expected when measurements access the correct $d$-dimensional quantum system and when the actual value of $\overline{\epsilon}$ is taken into account. \label{fig3}}
\end{figure}

We then proceed to test the NC inequality (\ref{IneKS21}) using as initial state each one of the 21~KS states $\ket{\text{KS}_i}$. For each initial KS state, a measurement run consisting of $1.0 \times 10^6$ faint pulses is performed. For each pulse, the random measurement projections are applied. Then we calculate the value of $\Sigma$ for that particular KS state from the recorded results. The final results are shown in Fig.~\ref{fig3}. We note that, for every initial state, the NC inequality is violated essentially by the same amount and that the observed violations are in very good agreement with the quantum prediction, even when experimental imperfections are taken into account.


\begin{figure}[tb]
\centering
\includegraphics[width=0.49 \textwidth]{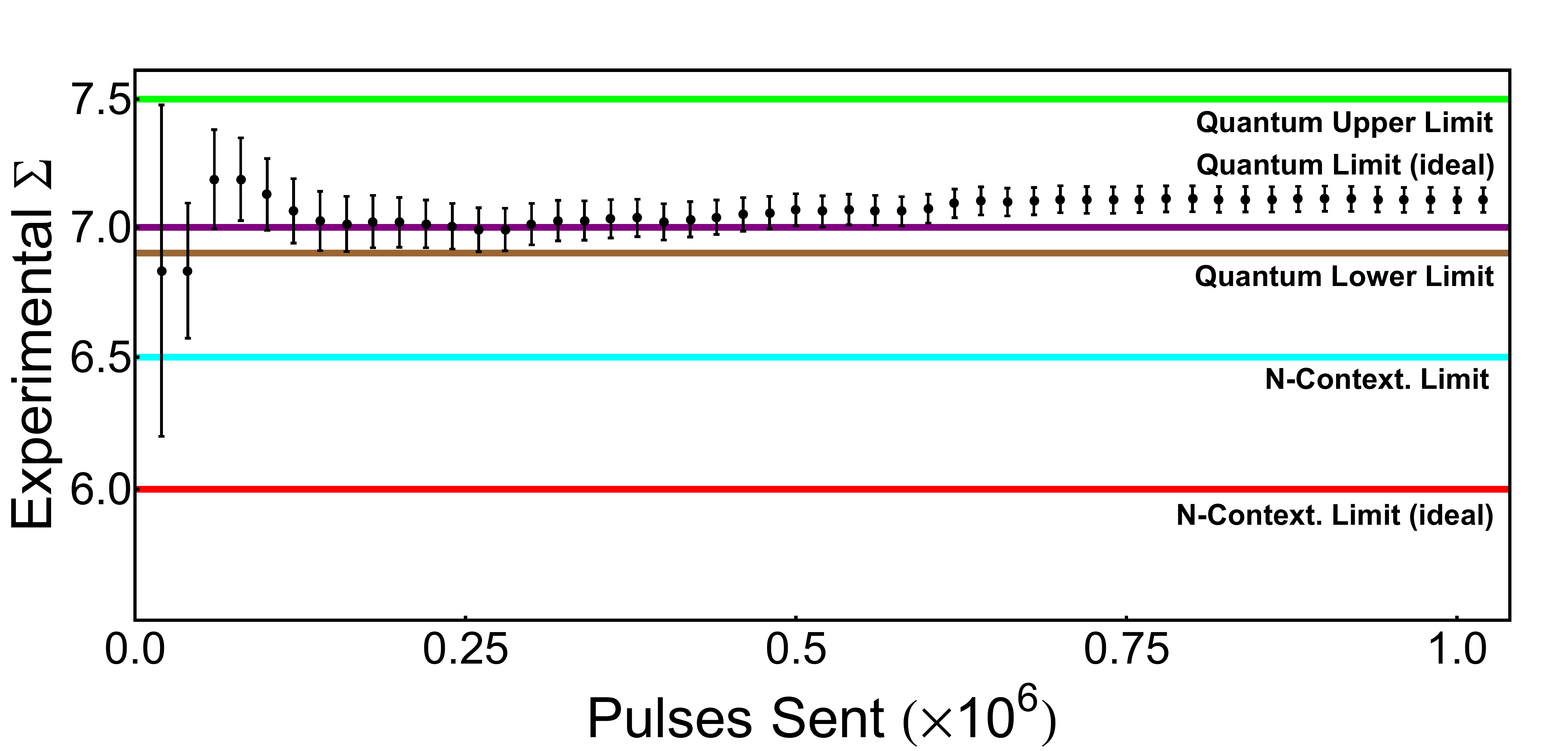}
\caption{(color online) Value of $\Sigma$, defined in Eq.~(\ref{IneKS21}), as a function of the number of pulses sent, while using $\ket{\text{KS}_{7}}=(0,0,0,0,0,1)$ as initial state.
\label{fig4}}
\end{figure}

An example of the measurement of $\Sigma$ as a function of the number of pulses sent is shown in Fig.~\ref{fig4}. It allows us to visualize how $\Sigma$ converges to its final value. The observed fluctuations in the beginning of the measurement procedure are mainly the result of statistical fluctuations. One can observe the convergence of the results as the statistical variations decrease when the number of detections increase with the elapsed time. The error bars are calculated considering a Poissonian distribution for the photon statistics.


\begin{figure}[tb]
\centering
\includegraphics[width=0.46 \textwidth]{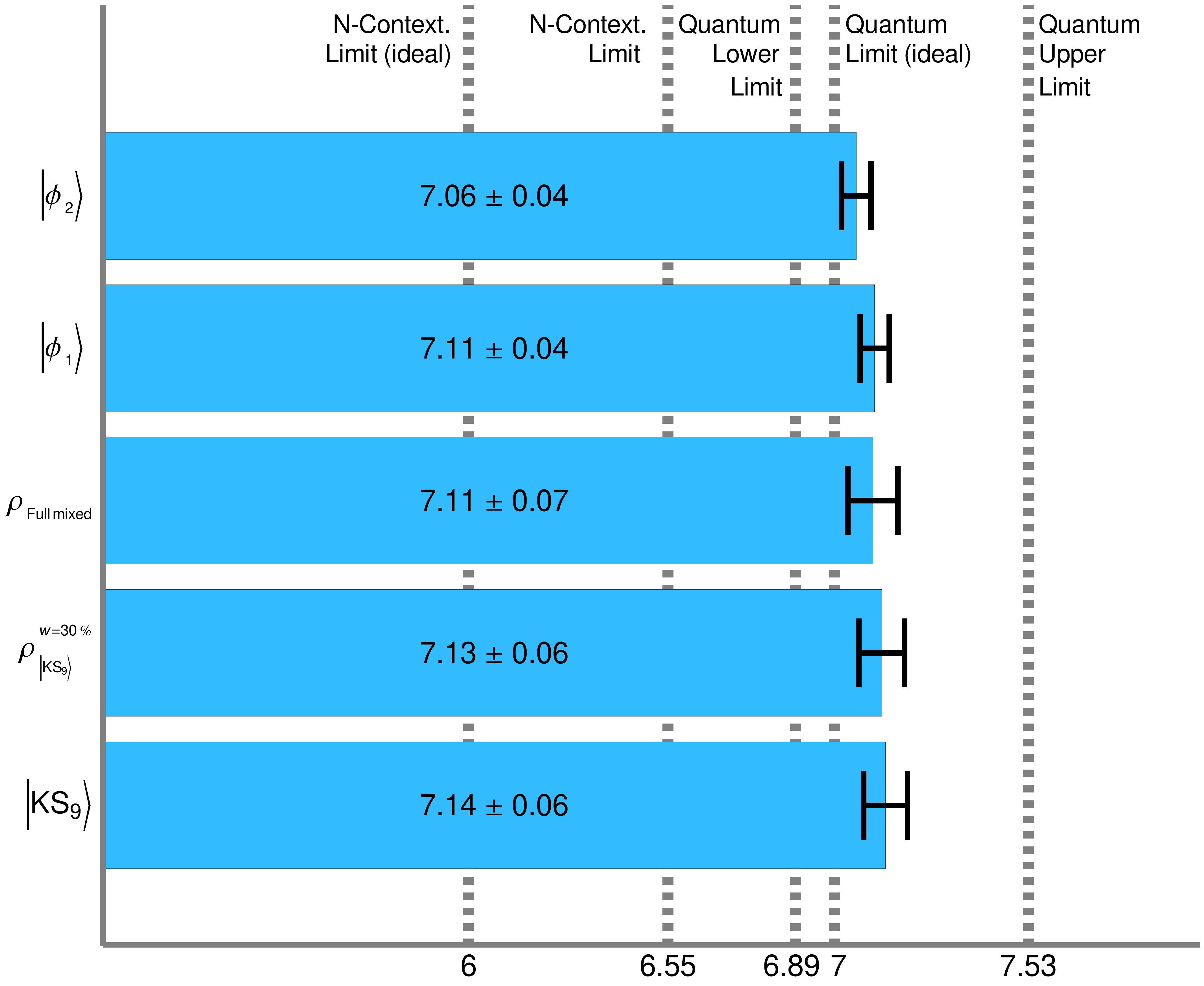}
\caption{(color online) Demonstration of the quantum state independency of the violation of inequality (\ref{IneKS21}) for the following states (normalization factors are omitted for simplicity): (i) $\ket{\phi_1} = (1,1,1,1,1,1)$, (ii) $\ket{\phi_2} = (1,0,0,0,1,0)$, (iii) a completely mixed state $\rho= \openone / 6$, (iv) $\rho_{\ket{\text{KS}_9}}^{w=30\%} = (1-w)\ket{\text{KS}_{9}}\bra{\text{KS}_{9}}+w \openone /6$, which is a partial mixture composed of the pure state $\ket{\text{KS}_{9}}= (0,1,0,1,\omega,\omega^2)$, where $\omega=e^{2 \pi i/3}$, and 30\% of added white noise, where $w$ is the amount of white noise and $\openone$ is the identity matrix in dimension 6, and (v) $\ket{\text{KS}_{9}}$. \label{fig5}}
\end{figure}


Finally, we complete the demonstration of the quantum state independency of the violation of inequality (\ref{IneKS21}) by testing $\Sigma$ for another five different initial states that do not belong to the KS set. The results of these tests are presented in Fig.~\ref{fig5}. Again, they show a clear violation of inequality (\ref{IneKS21}) in agreement with the quantum prediction, even when imperfections are taken into account. Notice that the observed value is essentially independent on the initial state of the system and the amount of white noise added.


{\em Conclusions.---}We have shown that, beyond their key role for proving fundamental results in quantum theory, KS sets can be extremely useful for practical quantum information processing involving high dimensional quantum systems. Specifically, by using six-dimensional quantum systems encoded in the transverse momentum of single photons, we have experimentally implemented for the first time the KS set that has the smallest number of contexts. We have used this KS set to illustrate a simple and efficient method to certify whether a set of measurements is actually accessing a previously established quantum six-dimensional system rather than a classical system or a different quantum system. This technique is of special relevance for quantum information processing due to the increasing complexity required for preparing and measuring quantum systems of high dimensions.


\begin{acknowledgments}
We thank C. Budroni, O. G\"uhne, M. Kleinmann, J.-{\AA}. Larsson, and G. Vallone for useful conversations. This work was supported by the Grants No.\ FONDECYT 1120067, No.\ CONICYT PFB08-024, No.\ Milenio P10-030-F (Chile), Project No.\ FIS2011-29400 (MINECO, Spain) with FEDER funds, and the FQXi large grant project ``The Nature of Information in Sequential Quantum Measurements.'' G.B.X. acknowledges financial support from FONDECYT 11110115. G.C., M.A. and E.S.G. acknowledge the financial support of CONICYT and AGCI.
\end{acknowledgments}



\end{document}